\documentstyle[12pt]{article}
\newif\ifhypertex
\ifx\hyperdef\UnDeFiNeD
    \hypertexfalse
    \def\hyperdef#1#2#3#4{#4}
    
    \def\e@tf@ur#1{}
    \def\hth/#1#2#3#4#5#6#7{{\tt hep-th/#1#2#3#4#5#6#7}}
    \def\CERN{CERN, Geneva, Switzerland}
\else
    \hypertextrue
  \def\hth/#1#2#3#4#5#6#7{
  {\tt hep-th/#1#2#3#4#5#6#7}}
\def\CERN{
CERN, Geneva, Switzerland}
\fi
\makeatletter
\@addtoreset{equation}{section}
\makeatother
\renewcommand{\baselinestretch}{1.1}

\newcommand{\equ}[1]{(\ref{#1})}
\newcommand{\be}{\begin{equation}}
\newcommand{\ee}{\end{equation}}
\newcommand{\eel}[1]{\label{#1}\end{equation}}
\newcommand{\bea}{\begin{eqnarray}}
\newcommand{\wha}{\widehat{\cal A}}
\newcommand{\eea}{\end{eqnarray}}
\newcommand{\eeal}[1]{\label{#1}\end{eqnarray}}
\newcommand{\beac}{\begin{equation}\begin{array}{rcl}}
\newcommand{\eeacn}[1]{\end{array}\label{#1}\end{equation}}

\newcommand{\del}{\partial}
\newcommand{\delb}{\bar{\partial}}
\newcommand{\eps}{\epsilon}

\newcommand{\ga}{\gamma}
\newcommand{\non}{\nonumber}

\newcommand{\QQ}{{\cal Q}}

\newcommand{\phb}{\bar{\varphi}}
\newcommand{\ph}{\varphi}
\newcommand{\psib}{{\bar{\psi}}}
\newcommand{\chib}{{\bar{\chi}}}
\newcommand{\mub}{{\bar{\mu}}}

\newcommand{\taub}{{\bar{\tau}}}

\begin{document}
%
\begin{titlepage}
\topskip0.5cm
\hfill\hbox{CERN-TH.7539/94}\\[-1.4cm]
\flushright{\hfill\hbox{hep-th/9412198}}\\[3.3cm]
\begin{center}{\Large\bf
Topological Strings from WZW Models\\[2cm]}{
\Large K. Landsteiner, W. Lerche and A. Sevrin\footnote{Permanent
address: Theoretische Natuurkunde, Vrije Universiteit Brussel, B-1050
Brussels,
Belgium.}
\\[1.2cm]}
\end{center}
\centerline{\CERN}
\vskip2.cm
\begin{abstract}
We show that the BRST structure of the topological string is encoded
in the ``small'' $N=4$ superconformal algebra, enabling us to obtain,
in a non-trivial way, the string theory from hamiltonian reduction of
$A(1|1)$. This leads to the important conclusion that not only
ordinary string theories, but topological strings as well, can
be obtained, or even defined, by hamiltonian reduction from WZW
models. Using two different gradations, we find either the standard
$N=2$ minimal models coupled to topological gravity, or an embedding
of the bosonic string into the topological string. We also comment
briefly on the generalization to super Lie algebras $A(n|n)$.
\end{abstract}
\vfill
\hbox{CERN-TH.7539/94}\hfill\\
\hbox{December 1994}\hfill\\
\end{titlepage}
%
%
\renewcommand{\baselinestretch}{1.1}\large\normalsize

It seems that the BRST structure of any string theory is encoded
in a (twisted) $N=2$ supersymmetric extension of its gauge algebra.
More precisely, the BRST structure of the bosonic string is
characterized by a twisted $N=2$ superconformal algebra \cite{bea},
that of the superstring by a twisted $N=3$ superconformal algebra
\cite{BLNW,BLLS}, that of $W_n$ strings by a twisted $N=2$ $W_n$
algebra \cite{BLNW,KI}, etc. The idea is that one can add the BRST
current and the anti-ghost to the gauge algebra, which then becomes
superconformally enlarged. The BRST-charge itself is then one of the
supercharges,
\bea
{{\cal Q}_{BRST}}\ \equiv\ G^+_0\ =\ \frac{1}{2\pi i}
\oint dz\,(cT+\dots)\ ,
\eeal{stringbrs}
while the anti-ghost $b(z)$ is the conjugate supercurrent, $G^-(z)$.
This automatically ensures that $T(z)=\{{{\cal Q}_{BRST}},b(z)\}$.

Though this structure is quite general, it becomes especially
important for non-critical strings. Here one can define the string
theory in an almost completely algebraic way through quantum
hamiltonian reduction \cite{HAMR} of an appropriate super-WZW model.
In order to fully characterize the reduction, one has to specify a
super-algebra and an embedding of $sl(2|1)$ into it. This already
uniquely determines the specific extended superconformal algebra.
Furthermore, one has to choose a particular gradation. This
determines the particular free-field realization, which must be such
that it allows for an interpretation in terms of string theory, ie.,
it must be of the form \equ{stringbrs}.

This approach of constructing string theories has the great advantage
that the calculations are of algorithmic nature, enabling one to
obtain the explicit form of the BRST operator in a relatively
straightforward manner (straightforward at least compared to the
usual trial and error method). Another advantage is that it also
systematically produces a consistent set of screening operators,
which are needed to properly define the free-field Hilbert space.

This program has been explicitly carried out for the non-critical
$W_n$ strings, based on a reduction of $sl(n|n-1)$ \cite{BLNW} and
strings with $N$ supersymmetries, based on a reduction of
$osp(N+2|2)$ \cite{BLLS}. The generalization to arbitrary embeddings
of $sl(2|1)$ in Lie super-algebras remains to be done, but we expect
it to yield a classification of at least a very large class of
non-critical string theories, if not of all of them. It has also
recently been shown that one can revert the reduction in that it is
possible to reconstruct the underlying Lie super algebra in terms of
the field content of a string theory \cite{SEMI}.

Up to now, however, it was not clear how to extend this program to
topological strings \cite{VVER}. The matter sector of a topological
string is made up by a particular realization of the twisted $N=2$
algebra with central charge $c_m$. This topological conformal field
theory is coupled to topological gravity, which can most easily
be represented by a supersymmetric ghost system consisting of
diffeomorphism ghosts, $b(z)$, $c(z)$, and their bosonic
superpartners, $\beta(z)$, $\ga(z)$ \cite{Dist}. Out of these ghost
systems, one can construct a twisted $N=2$ algebra with topological
central charge equal to $-9$. Since the ``true'' central charge of
each
building block vanishes identically and separately (as implied by the
twisting), the BRST-operator automatically squares to zero
independent from whether the {\it topological} central charges add up
to zero or not. Thus, there is no need to include an extra Liouville
sector\footnote{It has been shown \cite{VVER}, however, that the
inclusion of a Liouville sector makes the computation of physical
observables more feasible}. In that sense there is no ``critical''
central charge for topological strings. Nevertheless, it is
well-known that topological strings with $c_m=9$ have special
properties \cite{WBCOV}, and indeed, in our construction
given further below, this ``critical'' case of topological
strings will turn out to be distinguished.

In this letter we complete our program of obtaining all known string
theories from WZW models, by showing how topological strings can be
constructed from a suitable reduction of $sl(2|2)$. The possibility
of choosing different gradations will enable us to find two different
free-field realizations of the ``small'' $N=4$ algebra. Both have
interesting physical interpretations.

The first gradation results in $N=2$
minimal models coupled to topological gravity, represented by a
fermionic and a bosonic ghost-system, each with spins $(2,-1)$.

The second gradation results in an embedding of the non-critical
bosonic string into the topological string, in the sense that the
matter system is represented by the bosonic string. That is, we will
not only go one step further in the program initiated in \cite{BLLS}
of classifying string theories by purely Lie-algebraic methods, but
we also obtain a new mechanism of a string embedding that is defined
intrinsically in terms of hamiltonian reduction. This embedding looks
different from the kind of string embeddings developed in
\cite{BV,BO}, where one embeds string theories with $N$
supersymmetries into strings with $N+1$ supersymmetries; no obvious
connection to hamiltonian reduction could be made so far for these
embeddings.

Let us now describe the BRST-operator structure of topological
strings is some more detail. Neglecting the currents from the
Liouville sector, the total BRST charge, $Q_{BRST}=Q_s+Q_v$, consists
of the following two contributions \cite{VVER}:
\bea
Q_s&=&\oint\left( G_{+\, m}+G_{+\, gh}\right)\non\\
Q_v&=&\oint\, \left(c\,\left(T_m+\frac 1 2 T_{gh}\right)\,+ \,
\gamma\,\left(G_{-\,m}+\frac 1 2 G_{-\,gh}\right)\right)\ .
\eea
This implies that $T(z)$ is conjugate to $b(z)$ with respect to
$Q_v$: $T(z)\propto\{ Q_v,b(z)\}$, and to $G_-(z)$ with respect to
$Q_s$: $T(z)\propto\{Q_s,G_-(z)\}$, and suggests that the BRST
structure of the topological string should be related to a doubly
twisted, $N=4$ superconformal algebra \cite{LS,BLLS}.

There are essentially two $N=4$ algebras, a ``large'' one \cite{old},
which is obtained from the hamiltonian reduction of $D(2,1,\alpha)$,
and a ``small'' one \cite{ade}, obtained from the reduction of
$sl(2|2)$ (or $A(1|1)$ in a different notation) \cite{STT}. In
\cite{BLLS}, topological strings were related to a subalgebra of the
large $N=4$ algebra. However, no obvious connection with the
reduction of $D(2,1,\alpha)$ could be found. As we will now argue,
the more appropriate structure to look at is the small $N=4$ algebra,
which is generated by the energy-momentum tensor, four supercurrents
and an $su(2)$ affine Lie algebra.

Therefore our starting point is the Lie super algebra $sl(2|2)$. It
consists
of two $sl(2)$ affine algebras whose currents will be denoted by
$E^i(z)$ and $J^i(z)$,
with $i\in\{+,0,-\}$. The fermionic currents transform according to
the $(2|\bar{2})\oplus (\bar{2}|2)$ representation of the
$sl(2)\oplus sl(2)$ subalgebra. We thus denote the fermionic currents
by $j_{a}^{ij}(z)$, with $a,i,j \in \{+,-\}$. The
OPE's are explicitly given by\footnote{We used the metric
conventions $g^E_{00}=-g^J_{00}=-4$, $g^E_{+-}=-g^J_{+-}=-2$ and
$g^{+-}_{++--}=-2$, $g^{-+}_{+--+}=-2$, $g^{-+}_{++--}=2$,
$g^{+-}_{+--+}=2$.}
\bea
E^0 (z_1)\,E^0(z_2) = \frac{\kappa}{8}z_{12}^{-2}\, ,&\quad&
J^0 (z_1)\,J^0(z_2) = -\frac{\kappa}{8}z_{12}^{-2}\,,\non\\
E^0 (z_1)\,E^\pm(z_2) = \pm z_{12}^{-1}\frac{1}{2}\,E^\pm (z_2)\,
,&\quad&
J^0 (z_1)\,J^\pm(z_2) = \pm z_{12}^{-1}\frac{1}{2}\,J^\pm (z_2)\,,
\non\\
E^+ (z_1)\,E^-(z_2) = \frac{\kappa}{4}\,z_{12}^{-2} + z_{12}^{-1}E^0
(z_2)\,
,&\quad&
J^+ (z_1)J^-(z_2) = -\frac{\kappa}{4}z_{12}^{-2} + z_{12}^{-1}J^0
(z_2)\,,\non\\ E^0(z_1)j_a^{\pm i}(z_2) = \pm \frac{1}{4}z_{12}^{-1}
j_a^{\pm i}(z_2)\, ,
&\quad&
E^\pm(z_1)j_a^{\mp i}(z_2) = \frac{1}{2} z_{12}^{-1}j_a^{\pm
i}(z_2)\,,\non\\
J^0(z_1)j_a^{i\pm}(z_2) = \pm \frac{1}{4}z_{12}^{-1} j_a^{i
\pm}(z_2)\, ,&\quad&
J^\pm(z_1)j_a^{i\mp}(z_2) = \frac{1}{2}z_{12}^{-1}
j_a^{i\pm}(z_2)\,,\non
\eea
\bea
j_+^{\pm\pm}(z_1)j_-^{\mp\mp}(z_2) &=& \frac{\kappa}{4}z_{12}^{-2}
\pm
\frac{1}{2}z_{12}^{-1}\left( E^0(z_2)- J^0(z_2)\right)\,,\non\\
j_+^{\pm\mp}(z_1)j_-^{\mp\pm}(z_2) &=& -\frac{\kappa}{4}z_{12}^{-2}
\mp
\frac{1}{2}z_{12}^{-1}\left( E^0(z_2)+ J^0(z_2)\right)\,,\non
\eea
\bea
j_+^{\pm\pm}(z_1)j_-^{\pm\mp}(z_2) = -\frac{1}{2}z_{12}^{-1}E^\pm
(z_2), &\quad&
j_+^{\pm\mp}(z_1)j_-^{\pm\pm}(z_2) = \frac{1}{2}z_{12}^{-1}E^\pm
(z_2)\,,\non\\
j_+^{\pm\pm}(z_1)j_-^{\mp\pm}(z_2) =
\frac{1}{2}z_{12}^{-1}J^\pm (z_2), &\quad&
j_+^{\mp\pm}(z_1)j_-^{\pm\pm}(z_2) = -\frac{1}{2}z_{12}^{-1}J^\pm
(z_2)\,.
\eea

The relevant embedding of $sl(2|1)$ into $sl(2|2)$ is given by
choosing the bosonic generators of $sl(2|1)$ to be $E^i$ and $J^0$.
Note that we still have two possibilities for choosing the fermionic
part of the embedding. This traces back to the curious property of
$A(1|1)$ that the fermionic roots are simultaneously both, positive
and negative. We will see in the following that this does not really
constitute an ambiguity.

As was emphasized in \cite{BLLS}, the embedding just determines the
kind of superconformal algebra. However, in order to obtain a
specific free-field realization of this algebra that allows for a
string interpretation, one has to choose the gradation appropriately.
The
gradation will determine a consistent set of first class constraints
on the negatively graded part, and these constraints imply the
presence of a gauge symmetry. The gauge fixing and the construction
of the generators of the conformal algebra proceed then in a way
quite similar to what was developed in \cite{ST}. That is, although
the algebra is independent of the gradation, its realization, and in
particular the Miura transformation, depends on it.

We now apply this to $sl(2|2)$. The standard gradation, {\it i.e.}
the one used in \cite{ST}, is given by the eigenvalues of
$\mbox{ad}_{2 E^0}$. However, this gives rise to a ``symmetric''
free-field realization that has no interpretation as a string BRST
algebra. In order to find appropriate gradations, we note that in the
classification \cite{SLA} of Lie super-algebras one distinguishes
between algebras of type 2, whose fermionic part carries an
irreducible representation of the bosonic subalgebra, and algebras of
type 1, where this does not hold. Type 1 super Lie-algebras admit a
canonical gradation of the form $g=g_{-1/2}\oplus g_0\oplus g_{1/2}$,
where both $g_{1/2}$ and $g_{- 1/2}$ are irreducible representations
of $g_0$. In our case we have $j^{ij}_+ \in g_{+1/2}$ and $j^{ij}_-
\in g_{-1/2}$, while the remainder of the generators belong to $g_0$.

We find that we can use two different gradations, denoted by $I$ and
$II$, to describe sensible, string-type free-fields realizations.
Gradation $I$ is given by the sum of the canonical gradation with the
gradation defined by the action of $\mbox{ad}_{2(E^0+J^0)}$, whereas
gradation $II$ is defined by the eigenvalues of
$\mbox{ad}_{(2E^0+4J^0)}$ ignoring the canonical one. In the
following table, we explicitly give the corresponding grades for the
various elements of $sl(2|2)$:\\[1cm]
{\small \begin{tabular}{|c||c|c|c||c|c|c||c|c|c|c|c|c|c|c|}\hline
$ $&$E^+$&$E^0$&$E^-$&$J^+$&$J^0$&$J^-$&$j^{++}_+
$&$j^{+-}_+$&$j^{-+}_+$&$j^{--}_+$&$j^{++}_-$&$j^{+-}_-
$&$j^{-+}_-$&$j^{--}_-$\\ \hline
${\it I}$&$1$&$0$&$-1$&$1$&$0$&$-1$&$\frac 3 2
$&$\frac 1 2$&$\frac 1 2$&$-\frac{1}{2}$&$\frac 1
2$&$-\frac{1}{2}$&$-\frac{1}{2}$&
$-\frac{3}{2}$\\ \hline
${\it II}$&$1$&$0$&$-1$&$2$&$0$&$-2$&$\frac 3 2
$&$-\frac{1}{2}$&$\frac 1 2$&$-\frac{3}{2}$&$\frac 3 2$&
$-\frac{1}{2}$&$\frac{1}{2}$&
$-\frac{3}{2}$\\ \hline
\end{tabular}}\\[1cm]
Let us now first give a detailed discussion of the hamiltonian
reduction of $sl(2|2)$ using gradation $I$. The constraints we impose
on the negatively graded currents are of the form $\Phi^\alpha = 0$
with
\bea
\Phi^-_E = E^- - \frac{\kappa}{2},&\qquad&
\Phi^-_J = J^- - \mu,\non\\
\Phi^{+-}_- = j^{+-}_- - \psi,&\qquad&
\Phi^{--}_ += j^{--}_+ - \tau + \frac{1}{2} \psib\mu,\non\\
\Phi^{-+}_- = j^{-+}_- - \taub,&\qquad&
\Phi^{--}_- = j^{--}_- \ .
\eeal{constraintsI}
The auxiliary fields, $(\mu,\psi)$, and their conjugates,
$(\mub,\psib)$, were introduced in order to avoid that currents which
are in the kernel of $\mbox{ad}_{E^+}$ and would become the leading
term of the conformal currents, are constrained to zero (see
\cite{BLLS} for more details). The additional auxiliary fields
$(\tau,\taub)$ are needed to ensure that all constraints are first
class.\\ The constrained theory derives from the action\footnote{The
supertrace is defined by $str(xy) = x^a y^b g_{ab}(-1)^{deg(b)}$,
where $deg(b)=0$ or $1$, if it corresponds to a bosonic or fermionic
generator.}
\be
S_{inv} = \kappa S_{WZW}[g] + \frac{1}{\pi}\int\,
d^2z\,str(\bar{A}\Phi) - 	\frac{1}{2\pi}\int\, d^2z\,
(\psib\delb\psi
- \mub\delb\mu) - 	\frac{2}{\kappa\pi}\int\,d^2z\, \taub\delb\tau
\,,
\eel{invactionI}
where $S_{WZW}$ is the Wess-Zumino-Witten action on $SL(2|2)$ and the
gauge fields $\bar{A}$ appear as Lagrange multipliers that impose the
constraints. The gauge symmetry of the action is generated by the
strictly positively graded subalgebra of $sl(2|2)$:
\bea
\delta\bar{A} = \delb\eta + [\bar{A},\eta], &\quad& \delta J =
\del\eta + [J,\eta]\non\\
\delta\mu = 0, &\quad& \delta \mub = 2 \eta^+_J +
\eta^{++}_-\psib\non\\
\delta\psi = -\eta^{++}_-\mu &,& \delta\psib= -2\eta^{-+}_+\non\\
\delta\tau = \frac{\kappa}{2}\eta^{+-}_+, &\quad& \delta\taub =
\frac{\kappa}{2}\eta^{++}_-\,.
\eea
The kinetic terms of the auxiliary fields follow from requiring gauge
invariance and give rise to the following OPE's
\be
\psi(z_1).\psib(z_2) = \mu(z_1).\mub(z_2) = z_{12}^{-1}\quad
\tau(z_1).\taub(z_2) = \frac{\kappa}{4} z_{12}^{-1}\,.
\ee
We now quantize this theory by first fixing the gauge symmetry by
setting $\bar{A}=0$, and introducing the ghosts $C$ and anti-ghosts
$B$:
\bea
C &=& c_E^+ t^E_+ + c_J^+ t^J_+ + \ga^{-+}_+ t_{-+}^+ + \ga^{++}_-
t_{++}^- +
 \ga^{+-}_+ t_{+-}^+ + \ga^{++}_+ t_{++}^+\,\non\\
B &=& b^-_E t^J_- + b^-_J t^J_- + \beta^{+-}_- t_{+-}^- +
\beta^{--}_+ t_{--}^+ +
 \beta^{-+}_- t_{-+}^- + \beta^{--}_- t_{--}^-\,.
\eeal{ghosts}
The gauge fixing action reads \be
S_{g.f.} = \frac{1}{2\pi}\int
d^2z\, (D\bar{A} + B\delb C + \bar{A}\{B,C\}) \,,
\eel{Sgf}
where $D$ is a Langrange multiplier that imposes the gauge condition.
{}From this we can read off the OPE's for the ghosts
\bea
 b^-_E(z_1).c^+_E(z_2) = - b^-_J(z_1).c^+_J(z_2) =
-\beta^{+-}_+(z_1).\ga^{-+}_-(z_2)& = &\non\\
=-\beta^{--}_+(z_1).\ga^{++}_-(z_2) =
-\beta^{-+}_-(z_1).\ga^{+-}_+(z_2) =
\beta^{--}_-(z_1).\ga^{++}_+(z_2) &= &z_{12}^{-1}\,.
\eea
The BRST-charge then takes the standard form:
\bea
\QQ =\frac{1}{2\pi i}\oint str \left( C \Phi +\frac 1 2 C
J_{gh}\right),
\eea
where the ghost currents, $J_{gh} = \frac{1}{2}\{B,C\}$, are given
by:
\bea
E^+_{gh} &=& -\frac{1}{2} \beta^{+-}_-\ga^{++}_+,\qquad E^-_{gh} =
\frac{1}{2} \beta^{--}_-\ga^{-+}_+,\non\\ E^0_{gh} &=& -\frac{1}{2}
b^-_E c^+_E + \frac{1}{4} \beta^{+-}_-\ga^{-+}_+ -\frac{1}{4}
\beta^{--}_+\ga^{++}_- - \frac{1}{4} \beta^{-+}_-\ga^{+-}_+ +
\frac{1}{4} \beta^{--}_-\ga^{++}_+,\non\\ J^+_{gh} &=& -\frac{1}{2}
\beta^{-+}_-\ga^{++}_+,\qquad J^-_{gh} = \frac{1}{2}
\beta^{--}_-\ga^{+-}_+,\non\\ J^0_{gh} &=& \frac{1}{2} b^-_J c^+_J -
\frac{1}{4} \beta^{+-}_-\ga^{-+}_+ - \frac{1}{4}
\beta^{--}_+\ga^{++}_- + \frac{1}{4} \beta^{-+}_-\ga^{+-}_+ +
\frac{1}{4} \beta^{--}_-\ga^{++}_+,\non\\ j^{+-}_{+gh} &=&
-\frac{1}{2} b_J^-\ga^{++}_+ + \frac{1}{2} \beta^{--}_+ c^+_E,\qquad
j^{--}_{+gh} = \frac{1}{2} b^-_E\ga^{+-}_+ - \frac{1}{2}
b_J^-\ga^{-+}_+,\non\\ j^{-+}_{+gh} &=& \frac{1}{2} b_E^-\ga^{++}_+ -
\frac{1}{2} \beta^{--}_+ c^+_J,\qquad j^{++}_{-gh} = -\frac{1}{2}
\beta^{+-}_-c^+_J + \frac{1}{2} \beta^{-+}_- c^+_E,\non\\
j^{+-}_{-gh} &=& -\frac{1}{2} b_J^-\ga^{++}_- + \frac{1}{2}
\beta^{--}_- c^+_E,\qquad j^{-+}_{-gh} = \frac{1}{2} b_E^-\ga^{++}_-
- \frac{1}{2} \beta^{--}_- c^+_J.
\eeal{ghostcurrentsI}
We now proceed by closely following the methods of ref.\ \cite{ST}.
Classically, the generators of the extended conformal algebra are
given by the gauge invariant polynomials. At the quantum level this
translates to the fact the generators of the conformal algebra are
given by the generators of the cohomology of $\QQ$ on the algebra
$\cal A$, which consists of all normal ordered products of
$\{\hat{J}\equiv J + J_{gh},C,B,\tau,\taub,\psi,\psib,\mu,\mub\}$ and
their derivatives. The computation of the cohomology is not very hard
due to the presence of a double gradation of the complex ${\cal A}$:
\bea
{\cal A}=\bigoplus_{\stackrel{\scriptstyle m,n\in
\frac 1 2{\bf Z}}{m+n\in{\bf
Z}}}{\cal A}_{(m,n)}.
\eeal{complex}
The grade of the various fields is given by $(m,n)$, where $m$ is
the grade previously given in the table and $m+n$ is the ghost
number. The auxiliary fields have grade $(0,0)$. The BRST operator
splits into three parts, each of which has a definite grade:
\be
\QQ = \QQ_0 + \QQ_1 + \QQ_2\ ,
\ee
where
\bea
\QQ_0 &=& \frac{1}{2\pi i}\oint\frac{\kappa}{2} c^+_E -
 c^+_J\mu\,,\non\\
\QQ_1 &=&\frac{1}{2\pi i}\oint \ga^{-+}_+\psi +
\ga^{++}_-(\tau-\frac{1}{2}\psib\mu) + \ga^{+-}_+\tau\,,\non\\
\QQ_2 &=& \frac{1}{2\pi i}\oint-c^+_E(E^-+\frac{1}{2}E^-_{gh}) +
c^+_J(J^-+\frac{1}{2}J^-_{gh}) -
\ga^{-+}_+(j^{+-}_- + \frac{1}{2}j^{+-}_{-gh}) -\non\\
& &		-\ga^{++}_-(j^{--}_+ + \frac{1}{2}j^{--}_{+gh}) -
		\ga^{+-}_+(j^{-+}_- + \frac{1}{2}j^{-+}_{-gh}) +
\ga^{++}_+j^{--}_-,
\eea
which have grade
$(1,0)$, $(1/2,1/2)$ and
$(0,1)$, respectively.\\
Using the fact that $B$ and $\hat{\Phi}$ (where $\hat{\Phi}$ denotes
the substitution of $J$ by $\hat{J}$ in the constraints
\equ{constraintsI}) generate a sub-complex, one can argue along lines
similar to those in \cite{ST} that the cohomology of $\QQ$ on ${\cal
A}$ is isomorphic to the one computed on a reduced complex $\wha$.
The reduced complex is generated by those fields of ${\cal A}$ which
have grades $(m,n)$ with $m\geq 0$.\\ We thus introduce a
filtration $\wha^m$, $m\in\frac 1 2 {\bf Z}$ of $\wha$:
\be
\wha^m\equiv \bigoplus_{k\in\frac 1 2
{\bf Z}}\bigoplus_{l\geq m}\wha_{(k,l)}.
\ee
This leads to a spectral sequence $(E_r,d_r)$, $r\geq1$, converging
to $H^*(\wha;\QQ)$, where we have $E_r=H^*(E_{r-1};d_{r-1})$. The
sequence collapses already after the
second step: $E_1=H^*(\wha,\QQ_0)$, $E_2=E_\infty= H^*(E_1,\QQ_1)$,
where $E_2 \simeq H^*(\wha;\QQ)$ is generated by
$\{\hat{E}^+,\hat{J}^+,\hat{J}^0+\frac{1}{2}\mu\mub -
\frac{1}{4}\psib\psi -\frac{1}{\kappa}\tau\taub, \mu, \hat{j}^{++}_+,
\hat{j}^{++}_-, \hat{j}^{+-}_+ + \frac{2}{\kappa}\hat{j}^{-+}_+\mu,
\psi + \frac{2}{\kappa}\taub\mu\}$. The explicit form of the
generators now follows from a tic-tac-toe procedure and have the
structure:
\bea
\Lambda=\Lambda^{(m,-m)} + \Lambda^{(m-\frac{1}{2},-m+\frac{1}{2})}
+ ... + \Lambda^{(0,0)}
\eea
where the upper indices refer to the grades of the individual terms
and the leading terms $\Lambda^{(m,-m)}$ are the generators of $E_2$
given above. Using exactly the same arguments as those in \cite{ST},
one can show that $\Lambda\rightarrow \Lambda^{(0,0)}$ is an algebra
isomorphism, which is nothing but the quantum Miura transformation.
This provides us almost with the desired free-field realization. In
order to obtain the correct realization, we need in addition to
perform a similarity transformation (see also \cite{BLLS}) given by
$\Lambda^{(0,0)}{}' = S\, \Lambda^{(0,0)}\, S^{-1}$, where
$S = \exp \big[\frac{1}{\kappa\pi i} \oint\,dz\,  \mu \psib
\taub\big]$.
Bosonizing the Cartan currents
\bea
\hat{E}^0 = \frac{i\sqrt{\kappa}}{4} \del(\phb - \ph) ,\qquad
\hat{J}^0 = \frac{i\sqrt{\kappa}}{4} \del(\phb + \ph),\qquad
\ph(z_1).\phb(z_2) = ln(z_{12})\,,
\eea
and rescaling the auxiliary fields,
$\tau \rightarrow \frac{i\sqrt{\kappa}}{2}\tau,\qquad \taub
\rightarrow \frac{-i\sqrt{\kappa}}{2}\taub$,
then finally yields the following currents of the BRST algebra:
\bea
T &=& - \del\ph\del\phb + \frac{i\sqrt{\kappa}}{2}(\del^2\ph +
\del^2\phb)- 	\frac{1}{2}\tau\del\taub + \frac{1}{2}\del\tau\taub
+\non\\ & &	+\mu\del\mub - \frac{3}{2}\psi\del\psib -
\frac{1}{2}\del\psi\psib,\non\\ G_+ &=& \del\phb\taub + \psi\mub -
i\sqrt{\kappa}\del\taub,\non\\ G_- &=& -\del\ph\tau + \psib\del\mu +
2 \del\psib\mu + i \sqrt{\kappa}\del\tau,\non\\ \hat{G}_+ &=& \psib
(-\del\ph\del\phb - \tau\del\taub + i \sqrt{\kappa}\del^2\phb +
\mu\del\mub + \frac{1}{2} \del\mu\mub - \psi\del\psib)- \non\\ &
&-\mub ( - \del\phb\tau + i\sqrt{\kappa}\del\tau +
\del\psib\mu + \frac{1}{2}\psib\del\mu) - \non\\ &
&-\del(\psib(i\sqrt{\kappa}\del\phb - i\sqrt{\kappa}\del\ph -
\tau\taub + \mu\mub)) - (1+\kappa) \del^2\psib,\non\\ \hat{G}_- &=&
\psi,\non\\ K_+ &=& -\psib\taub\del\phb +
i\sqrt{\kappa}\psib\del\taub + \mub\psi\psib - \mub^2\mu
+\mub\tau\taub - i \sqrt{\kappa}\mub\del\phb +\non\\ & & + i
\sqrt{\kappa}\mub\del\ph - (1+\kappa)\del\mub ,\non\\ K_3 &=& i
\sqrt{\kappa}(\del\phb-\del\ph) + 2\mu\mub - \psi\psib -
\tau\taub,\non\\ K_- &=& \mu.
\eea
They are canonically normalized such that they generate the small
$N=4$ algebra:
\bea
G_+(z_1)G_-(z_2) &=& \hat{G}_+(z_1)\hat{G}_-(z_2) =
\frac{c}{3}z_{12}^{-3} + z_{12}^{-2}K_3(z_2) + z_{12}^{-1}
(T+\frac{1}{2}\del K_3)(z_2),\non\\ G_{\pm}(z_1)\hat{G}_{\pm}(z_2)
&=& \mp 2 z_{12}^{-2}K_{\pm}\mp z_{12}^{-1}\del K_{\pm}(z_2),\non\\
K_{\pm}(z_1)G_{\mp}(z_2) &=& \pm z_{12}^{-1}\hat{G}_{\pm}(z_2),\qquad
K_{\pm}(z_1)\hat{G}_{\mp}(z_2) = \mp z_{12}^{-1} G_{\pm}(z_2),\non\\
K_3(z_1)G_{\pm}(z_2) &=& \pm z_{12}^{-1} G_{\pm}(z_2),\qquad
K_3(z_1)\hat{G}_{\pm}(z_2) = \pm z_{12}^{-1}\hat{G}_{\pm}(z_2)\non\\
K_l(z_1)K_m(z_2) &=& \frac{c}{6}\delta_{lm}z_{12}^{-2}+ \eps_{lm}{}^n
z_{12}^{-1}K_n(z_2),
\eea
where $c=-6(1+\kappa )$.

The interpretation of the above formulas in terms of string theory
evident: the bosons $\ph$, $\phb$ together with the fermions $\tau$,
$\taub$ represent a minimal $N=2$ matter free-field realization, with
background charge given by $i \sqrt{\kappa}$. Furthermore, the string
ghost-antighost pair corresponds to $\psib$, $\psi$, whereas $\mub$,
$\mu$ represents the bosonic ghost-antighost pair of the topological
string. After twisting the energy-momentum tensor $T\rightarrow
T+\frac{1}{2}\del K_3$, $\psi$ and $\mu$ acquire spin 2 and $\psib$
and $\mub$ spin $-1$.

The specific realization of the $N=2$ sub-system that we interpret
here
as matter system is not essential for the closure of the algebra. In
fact, it is easy to see that we can replace it by any other
realization of the $N=2$ algebra. In contrast to the stringy
hamiltonian reductions considered so far, we do not get a Liouville
system. This seems to be natural, since as mentioned in the
introduction, one would not expect the Liouville field to play an
essential role in topological gravity. Let us also note that the only
physical observable in equivariant cohomology, namely the generator
of gravitational descendents, $\del\mub$, appears here naturally as
part of the $sl(2)$ Kac-Moody current $K_+$. The central extension of
this algebra vanishes and $\del\mub$ decouples from the algebra
precisely if $\kappa=-1$ which constitutes the ``critical case''
$c_m=9$ of topological strings.

The screening operators associated with this free-field
realization are easily obtained from $sl(2|2)$. The following
screeners are related to the three simple fermionic roots:
\bea
S_1 &=& \frac{1}{2\pi i} \oint\,dz\, \taub
\exp(-\frac{i}{\sqrt{\kappa}}\ph)\,\non\\ S_2 &=& \frac{1}{2\pi i}
\oint\,dz\, \tau \exp(-\frac{i}{\sqrt{\kappa}}\phb)\,\non\\ S_3 &=&
\frac{1}{2\pi i} \oint\,dz\, (\psi + \frac{i}{\sqrt{\kappa}}\mu\taub)
\exp(-\frac{i}{\sqrt{\kappa}}\ph)\,.
\eea
They are needed to define the physical Hilbert
space of the theory.

We now turn to discussing the reduction for gradation $II$.
Here, the first-class constraints look
\bea
\Phi^-_E = E^- - \frac{\kappa}{2}+\frac 1 4 \psib\chib\mu,&\qquad&
\Phi^-_J = J^- - \mu,\non\\
\Phi^{+-}_+ = j^{+-}_+ - \psi,&\qquad&
\Phi^{+-}_- = j^{+-}_- - \chi,\non\\
\Phi^{--}_+ = j^{--}_+ + \frac 1 2 \mu\chib,&\qquad&
\Phi^{--}_- = j^{--}_- - \frac 1 2 \mu\psib\ .
\eeal{constraintsII}
The action of the constrained theory is now given by
\be
S_{inv} = \kappa S_{WZW}[g] + \frac{1}{\pi}\int\,
d^2z\,str(\bar{A}\Phi) - 	\frac{1}{2\pi}\int\, d^2z\,
(\psib\delb\psi
+ \chib\delb\chi - \mub\delb\mu)  \,,
\eel{invactionII}
and the OPE's for the auxiliary fields are
$\psi(z_1).\psib(z_2) = \chi(z_1).\chib(z_2) = \mu(z_1).\mub(z_2) =
z_{12}^{-1}$.
The action is invariant under the following gauge transformations:
\bea
\delta\bar{A} = \delb\eta + [\bar{A},\eta], &\quad& \delta J =
\del\eta + [J,\eta]\non\\
\delta\mu = 0, &\quad& \delta \mub = 2 \eta^+_J + \chib\eta^{++}_- +
 \psib\eta^{++}_+ + \frac 1 2 \eta_E^+\psib\chib\\
\delta\psi =  \eta^{++}_+\mu + \frac 1 2 \eta^+_E\chib\mu, &\quad&
\delta\psib= -2\eta^{-+}_-\non\\
\delta\chi =  \eta^{++}_-\mu - \frac 1 2 \eta^+_E\psib\mu, &\quad&
\delta\chib= 2\eta^{-+}_+\,.
\eea
To quantize the theory, we can use almost the same ghost system as in
\equ{ghosts}, except that we have to replace the
$(\beta^{-+}_-,\gamma^{+-}_+)$-system by
$(\beta^{+-}_-,\gamma^{-+}_+)$,
which obeys:
$\beta^{+-}_-(z_1),\gamma^{-+}_+(z_2) = - z_{12}^{-1}$.
The ghost currents $j^{+-}_+\,,j^{-+}_+\,,j^{+-}_-\,,j^{-+}_-$ stay
the same as before and therefore can directly be taken over from
\equ{ghostcurrentsI}. The other ghost currents get changed, however
(notice, e.g., that this time there are no ghost contributions to
$J^\pm$):
\bea
E^+_{gh} &=& \frac{1}{2} \beta^{+-}_+\ga^{++}_- - \frac{1}{2}
\beta^{+-}_-\ga^{++}_+,
\qquad
 E^-_{gh} = -\frac{1}{2} \beta^{--}_+\ga^{-+}_- + \frac{1}{2}
\beta^{--}_-\ga^{-+}_+,\non\\
E^0_{gh} &=& -\frac{1}{2}b^-_E c^+_E - \frac{1}{4}
\beta^{+-}_+\ga^{-+}_- -\frac{1}{4}\beta^{--}_+\ga^{++}_- +
\frac{1}{4} \beta^{+-}_-\ga^{-+}_+ +
\frac{1}{4} \beta^{--}_-\ga^{++}_+,\non\\
J^0_{gh} &=& \frac{1}{2} b^-_J c^+_J +
\frac{1}{4} \beta^{+-}_+\ga^{-+}_- - \frac{1}{4}
\beta^{--}_+\ga^{++}_- - \frac{1}{4} \beta^{+-}_-\ga^{-+}_+ +
\frac{1}{4} \beta^{--}_-\ga^{++}_+,\non\\
j^{++}_{+gh} &=& -\frac 1 2 \beta^{+-}_- c^+_J\qquad
j^{--}_{+gh} = - \frac{1}{2}b_J^-\ga^{-+}_+,\non\\
j^{++}_{-gh} &=& -\frac{1}{2}
\beta^{+-}_-c^+_J, \qquad
j^{--}_{-gh} = - \frac{1}{2} b^{-}_J \ga^{-+}_-.
\eea
The most drastic difference in comparison to gradation $I$ shows up
in the structure of the BRST operator. Using the new gradation to
define the complex \equ{complex}, the BRST charge splits into the
following five parts:
\bea
\QQ_0 &=& \frac{1}{2\pi i}\oint \left( -
 c^+_J\mu \right)\,,\qquad
\QQ_1 = \frac{1}{2\pi i}\oint\,\left(-\frac{1}{2} \ga^{++}_-\chib\mu
-
 \frac{1}{2} \ga^{++}_+\psib\mu\right)\,,\non\\
\QQ_2 &=& \frac{1}{2\pi i}\oint \left( \frac \kappa 2 c^+_E - \frac 1
4
c^+_E\psib\chib\mu \right)\,,\qquad
\QQ_3 =\frac{1}{2\pi i}\oint \left(-\ga^{-+}_-\psi + \ga^{-+}_+
\chi\right)\,,\non\\
\QQ_4 &=& \frac{1}{2\pi i}\oint\left(-c^+_E(E^-+\frac{1}{2}E^-_{gh})
+ c^+_J J^- +
\ga^{-+}_-(j^{+-}_+ + \frac{1}{2}j^{+-}_{+gh}) -\right.\non\\
& &\left.-\ga^{++}_-(j^{--}_+ + \frac{1}{2}j^{--}_{+gh}) -
\ga^{-+}_+(j^{+-}_- + \frac{1}{2}j^{+-}_{-gh}) -
\ga^{++}_+(j^{--}_- + j^{--}_{-gh})\right),
\eea
They individually square to zero, and their bigrades are given by
$(2,-1)$,
$(\frac 3 2 ,\frac 1 2 )$, $(1,0)$, $(\frac 1 2 , \frac 1 2 )$ and
$(0,1)$. Moreover, they obey the relations
$\{\QQ_1,\QQ_3\} + \{\QQ_0,\QQ_4\} =
\{\QQ_2,\QQ_3\} + \{\QQ_1,\QQ_4\} = 0\,$.
The anti-ghosts $B$ and the constraints $\hat{\Phi}$ form
BRST-doublets and we can define a reduced complex, completely
parallel to the treatment of gradation $I$. To compute the cohomology
on this reduced complex, we note that we can write the BRST-operator
as $\QQ_0 + \QQ^\prime$ with $\QQ^\prime = \sum^{4}_{i=1}\QQ_i$. Then
this defines a double complex in the weak sense of \cite{ST}. Of
course, in the spectral sequence we thus have $d_1 = Q^\prime$. But
now we can use the splitting given above, such that $Q^\prime=Q_1+
Q^{\prime\prime}$ and $Q^{\prime\prime} = \sum^{4}_{i=2}\QQ_i$.
Proceeding in this manner, it is straightforward to compute the
cohomology in a stepwise fashion. Since we use here a rather
non-standard decomposition of the BRST-operator, it is useful to
denote the non-trivial action of the relevant parts of it on the
fields explicitly:
\bea
\QQ_0 :& &[\QQ_0,J^0] = \frac 1 2 c^+_J\mu,\qquad [\QQ_0,\mub] =
-c_J^+ \,,\non\\
\QQ_1 :& &[\QQ_1,\mub] = -\frac 1 2 \ga^{++}_-\mu\chib - \frac 1 2
\ga^{++}_+
\psib,\qquad \{\QQ_1,\psi\} = -\frac 1 2 \ga^{++}_+\mu, \qquad
\{\QQ_1,\chi\} =-\frac 1 2 \ga^{++}_-\mu,\non\\
& &[\QQ_1,\hat{E}^0] = \frac 1 8 \ga^{++}_-\chib\mu +
\frac 1 8 \ga^{++}_+\psib\mu,\qquad
[\QQ_1,\hat{J}^0] = \frac 1 8 \ga^{++}_-\chib\mu +
\frac 1 8 \ga^{++}_+\psib\mu,\non\\
& &\{\QQ_1,\hat{j}^{-+}_+\} = -\frac 1 4 c^+_J\chib\mu,\qquad
\{\QQ_1,\hat{j}^{-+}_-\} =  \frac 1 4 c^+_J\chib\mu,\non\\
\QQ_2 :& &[\QQ_2,\mub] = -\frac 1 4 c^+_E\psib\chib,\qquad
\{\QQ_2,\psi\} = \frac 1 4 c^+_E\chib\mu,\qquad
\{\QQ_2,\chi\} = -\frac 1 4 c^+_E\psib \mu,\non\\
& &[\QQ_2,\hat{E}^0] = - \frac{\kappa}{4} c^+_E - \frac 1 8
c^+_E\chib\psib\mu,\qquad
\{\QQ_2,\hat{j}^{-+}_+\} = \frac \kappa 4 \ga^{++}_+ +
\frac 1 8 \ga^{++}_+\chib\psib\mu,\non\\
& &\{\QQ_2,\hat{j}^{-+}_-\} = \frac \kappa 4 \ga^{++}_- + \frac 1 8
\ga^{++}_-
\chib\psib\mu,\non\\
\QQ_3 :& &\{\QQ_3,\psib\} = - \ga^{-+}_-,\qquad\{\QQ_3,\chib\} =
\ga^{-+}_+, \qquad
\{\QQ_3,\hat{j}^{++}_+\} = \frac 1 2 c^+_J\psi,\non\\
& &\{\QQ_3,\hat{j}^{++}_-\} = \frac 1 2 c^+_J\chi,\qquad
[\QQ_3,\hat{E}^+] = \frac 1 2 \ga^{++}_-\psi - \frac 1 2
\ga^{++}_+\chi,\non\\
& &[\QQ_3,\hat{E}^0] = -\frac 1 4 \ga^{-+}_-\psi + \frac 1 4
\ga^{-+}_+\chi,\qquad
[\QQ_3,\hat{J}^0] = \frac 1 4 \ga^{-+}_-\psi - \frac 1 4
\ga^{-+}_+\chi\,.
\eeal{qaction}
{}From these equations we immediately see that $(\mub,c^+_J)$
decouple from the $\QQ_0$-cohomology. Then, looking at $\QQ_1$, we
note that the auxiliary fields $(\psi,\chi)$ lie in BRST-doublets
together with the composite operators
$(\ga^{++}_+\mu,\ga^{++}_-\mu)$. This means that these decouple from
the $\QQ_1$-cohomology. In the next step we have to take into account
that in the complex $E_2 = H^*(\hat{{\cal A}};\QQ_1)$ there are
vanishing relations of the form $\ga^{++}_+\mu = \ga^{++}_-\mu \sim
0$. Therefore we find that the composite fields $\hat{j}^{-+}_+\mu$
and $\hat{j}^{-+}_-\mu$ are non-trivial elements of the
$\QQ_2$-cohomology on $E_2$. Proceeding further one finds that the
spectral sequence collapses after the fourth step and that the
cohomology of $\QQ$ is spanned by
$\{\hat{E}^+,\hat{J}^+,\hat{J}^0+\frac 1 2 \mu\mub - \frac 1 4
\psi\psib - \frac 1 4 \chi\chib,\hat{j}^{++}_+,\hat{j}^{++}_-,
\hat{j}^{-+}_+\mu,\hat{j}^{-+}_-\mu,\mu\}$. It is surprising that not
all the elements that lie in $ker(ad_{E^+})$ correspond to leading
terms in the cohomology. It is however not difficult to see, by
following a generalized tic-tac-toe procedure, that one picks up the
highest weight auxiliaries $\psi$, $\chi$ in the tail of
$\hat{j}^{-+}_+\mu$ and $\hat{j}^{-+}_-\mu$. It is also clear that
the truncation to zero grade fields still provides an algebra
isomorphism.

In order to end up with a string-type free-field
realization, we demand that one of the supercharges is given
by just a single auxiliary field. We could choose $\chi$, or, equally
well, $\psi$. To be specific, let us choose $\chi$, and perform the
following similarity transformation:
\be
S = \exp \Big[-\frac{1}{2\pi i} \oint\,dz \,( \frac 1 \kappa
\chib\hat{E}^0\psib\mu + \frac 1 2 \chib\del\psib\mu + \frac 1
\kappa\chib\hat{J}^0\psib\mu)\Big].
\ee
Furthermore, we bosonize the Cartan currents:
\be
\hat{E}^0 = \frac{i\sqrt{\kappa}}{2\sqrt{2}}\del\ph_1,\qquad
\hat{J}^0 = \frac{\sqrt{\kappa}}{2\sqrt{2}}\del\ph_2,\qquad
\ph_i(z_1).\ph_j(z_2) = - \delta_{ij} \,ln (z_{12})
\ee
Then, finally, we arrive at a second free-field realization of the
small
$N=4$ algebra that can attributed to topological string theory:
\bea
T &=& - \frac 1 2 (\del\ph_1)^2 +
\frac{i(1-\kappa)}{\sqrt{2\kappa}}\del^2\ph_1 - \frac 1 2
(\del\ph_2)^2 + \frac{1}{\sqrt{2\kappa}}\del^2\ph_2 + \non\\
& &+\mu\del\mub - \frac 3 2 \chi\del\chib - \frac 1 2 \del\chi\chib -
\frac 3 2 \psi\del\psib -
\frac 1 2 \del \psi\psib \non\\
G_+ &=& -\psib(- \frac 1 2 (\del\ph_1)^2 +
\frac{i(1-\kappa)}{\sqrt{2\kappa}}\del^2\ph_1 - \frac 1 2
(\del\ph_2)^2 + \frac{1+\kappa}{\sqrt{2\kappa}}\del^2\ph_2 -
\psi\del\psib) +\non\\ & &+\sqrt{2\kappa}\del(\psib\del\ph_1) -
\frac{1-2\kappa}{2}\del^2\psib +\chi\mub\non\\
G_- &=& -\psi - \chib\del\mu -2 \del\chib\mu\non\\
\hat{G}_+ &=& \chib[- \frac 1 2 (\del\ph_1)^2 +
\frac{i(1-\kappa)}{\sqrt{2\kappa}}\del^2\ph_1 - \frac 1 2
(\del\ph_2)^2 + \frac{1+\kappa}{\sqrt{2\kappa}}\del^2\ph_2 + \non\\
& &- 2 \psi\del\psib - \del\psi\psib + \mu\del\mub + \frac 1 2
\del\mu\mub - \chi\del\chib] - \non\\
& &-\mub(\psi + \frac 1 2 \chib\del\mu + \del\chib\mu) -
\del[\chib(\sqrt{2\kappa}\del\ph_2) - \psi\psib + \mu\mub ] -
(1+\kappa)\del^2\chib\non\\
\hat{G}_- &=& \chi\non\\
K_+ &=& \frac 1 2 \chib\psib(\del\ph_1)^2 + \frac 1 2
\chib\psib(\del\ph_2)^2 -
\frac{i(1-\kappa)}{\sqrt{2\kappa}}\chib\psib\del^2\ph_1
-\frac{1-\kappa}{\sqrt{2\kappa}}\chib\psib\del^2\ph_2 -
\chib\del\psib\del\ph_2 +  \non\\
& &+\mub\chi\chib - \mub^2\mu - \sqrt{2\kappa}\mub\del\ph_2 +
\mub\psi\psib + \chib\psib\psi\del\psib -
\frac{1-\kappa}{2}\chib\del^2\psib - (1+\kappa)\del\mub\non\\
K_3 &=& \sqrt{2\kappa}\del\ph_2 - \chi\chib - \psi\psib + 2
\mu\mub\non\\
K_- &=& \mu
\eea
After twisting, the auxiliary fields, $(\psi,\chi,\mu)$ have spin $2$
and $(\psib,\chib,\mub)$ have spin $-1$. From the form of the
generators we see that we can indeed interpret $(\chi,\chib)$ and
$(\mu,\mub)$ as the ghosts of the topological string. The rest
constitutes the matter sector of the topological string and is
isomorphic to a particular realization of the ordinary bosonic
string! That is, $\ph_1$ is the matter and $\ph_2$ the Liouville
system of the bosonic string, and the diffeomorphism ghosts are given
by $(\psi,\psib)$. Indeed, modulo the terms arising from the
topological ghost-system, the currents $T$, $G_+$, $G_-$ and $K_3$
are exactly the well-known $N=2$-currents of the bosonic string
\cite{bea,BLNW}. This means that the above formulas should be
interpreted as an embedding of the non-critical bosonic string into
the topological string. Note that the critical topological string,
which corresponds to $\kappa=-1$, coincides with the critical bosonic
string, as can be seen by the vanishing of the background charges.

For sake of completeness, we present the following screening
operators,
\bea
S_1 &=& \frac{1}{2\pi i} \oint\,dz\, \exp(-i\sqrt{2\kappa} \ph_1
)\non\\
S_2 &=& \frac{1}{2\pi i} \oint\,dz\, \psi
\exp[-\frac{i}{\sqrt{2\kappa}} (\ph_1-i\ph_2) ]\non\\
S_3 &=& \frac{1}{2\pi i} \oint\,dz\, (\chi + 2 \mu\del\psib
+\del\mu\psib) \exp[\frac{-i}{\sqrt{2\kappa}}(\ph_1-i\ph_2)]\,,
\eea
that are necessary to properly define the free-field Hilbert space.

Generalizing the above considerations to topological
$W_n$-strings \cite{LS}, one expects that a reduction of $sl(n|n)$
should be relevant. Let us consider the reduction of $sl(3|3)$ in
some more detail. As we want to have an $N=2$ $W_3$ algebra as a part
of the full string BRST algebra, we need to analyze the embedding of
$sl(3|2)$ into $sl(3|3)$. The embedding of $sl(2|1)$ in $sl(3|2)$ is
then precisely the one that gives rise to the $N=2$ $W_3$ algebra. In
terms of $sl(2)$ representations, we find that the bosonic part of
$sl(3|3)$ will give rise to a spin $3$, two spin $2$, two spin
$\frac{3}{2}$ and one spin $1$ current. The fermionic part yields two
sets of currents with spins $\frac{5}{2}$, $2$ and $\frac{3}{2}$.
Thus we find 6 bosonic and 6 fermionic generators. However, from a
$W_3$ extension of the small $N=4$ algebra one expects $8$ bosonic
and $8$ fermionic generators. We thus see from this simple counting
argument that the reduction of $sl(3|3)$ will not give rise to a
naive formulation of topological $W_3$-gravity \cite{LS}.

It is nevertheless worth trying to go on with the analysis. It is
easy to find a gradation analogous to the one we used for $sl(2|2)$,
where two of the $sl(2)$ highest weights, one belonging to a
fermionic triplet and one belonging to a bosonic doublet, have
negative grades. {}From our discussion of $sl(2|2)$ it is clear that
this will result in bosonic and fermionic ghost systems. Under the
$U(1)$-current of $sl(3|3)$ (the one which corresponds to the ghost
current after the reduction), these two ghosts systems have charges
$1$ and $\frac{3}{2}$, respectively, such that both would obtain spin
3 in the twisted superconformal algebra. They should therefore
represent the high-spin ghost-systems of topological $W_3$-gravity.
We thus expect that one obtains from such a reduction topological
$W_3$-gravity to be realized in a kind of ``half-rotated'' matter
picture, where the low-spin ghosts are decoupled from the theory.
That such a realization exists is in fact quite plausible since we
know that the spectrum of topological $W_3$-gravity can be
represented entirely in the matter sector \cite{LS}.

Combining the results of this paper with \cite{BLLS}, we are drawn to
conclude that a very systematic and almost completely algebraic
approach to string theory might be attainable, at least as far as
controlling and classifying the possible gauge structures on the
world-sheet goes. Indeed, these two papers demonstrate that
all known string theories can be obtained in a
straightforward way from reducing WZW models. As all simple
supergroups have been classified, the natural step to take now is a
complete classification of string theories that can be obtained this
way. This is presently under investigation \cite{RSS}. We hope that
such a group theoretical approach to string theory will eventually
provide clues to the structure of a ``universal'' string field theory
that would have vacua corresponding to arbitrary gauge groups on the
world-sheet.

%
%
\noindent {\bf Acknowledgements}
We extensively used Kris Thielemans Mathematica package OPEdefs.m
\cite{OPEdefs} for the computations.
%
%

\end{document}